# MRI Reconstruction Using Deep Energy-Based Model

Yu Guan[1], Zongjiang Tu[1], Shanshan Wang[2], Qiegen Liu[1], Yuhao Wang[1] and Dong Liang[2,3]

[1] *Department of Electronic Information Engineering, Nanchang University, Nanchang 330031, China.*

[2] *Paul C. Lauterbur Research Center for Biomedical Imaging, Shenzhen Institutes of Advanced Technology, Chinese Academy of Sciences, Shenzhen 518055, China.*

[3] *Medical AI research center, Shenzhen Institutes of Advanced Technology, Chinese Academy of Sciences, Shenzhen 518055, China.*

***Correspondence to:**

Dong Liang, Ph.D.

Paul C. Lauterbur Research Centre for Biomedical Imaging

Institute of Biomedical and Health Engineering

Shenzhen Institutes of Advanced Technology

Chinese Academy of Sciences, Shenzhen, Guangdong, China, 518055

Tel: (86) 755-86392243

Fax: (86) 755-86392299

Email: dong.liang@siat.ac.cn

# Abstract

**Purpose:** Although recent deep energy-based generative models (EBMs) have shown encouraging results in many image generation tasks, how to take advantage of the self-adversarial cogitation in deep EBMs to boost the performance of Magnetic Resonance Imaging (MRI) reconstruction is still desired.

**Methods:** With the successful application of deep learning in a wide range of MRI reconstruction, a line of emerging research involves formulating an optimization-based reconstruction method in the space of a generative model. Leveraging this, a novel regularization strategy is introduced in this article which takes advantage of self-adversarial cogitation of the deep energy-based model. More precisely, we advocate for alternative learning a more powerful energy-based model with maximum likelihood estimation to obtain the deep energy-based information, represented as image prior. Simultaneously, implicit inference with Langevin dynamics is a unique property of reconstruction. In contrast to other generative models for reconstruction, the proposed method utilizes deep energy-based information as the image prior in reconstruction to improve the quality of image.

**Results:** Experiment results that imply the proposed technique can obtain remarkable performance in terms of high reconstruction accuracy that is competitive with state-of-the-art methods, and does not suffer from mode collapse.

**Conclusion:** Algorithmically, an iterative approach was presented to strengthen EBM training with the gradient of energy network. The robustness and the reproducibility of the algorithm were also experimentally validated. More importantly, the proposed reconstruction framework can be generalized for most MRI reconstruction scenarios.

## Introduction

Magnetic Resonance Imaging (MRI) is a sophisticated and versatile medical imaging modality for numerous clinical applications. Furthermore, it is non-intrusive, high-resolution, and safe to living organisms. Even though MRI does not use dangerous radiation for imaging, its long acquisition time causes discomfort to patients and hinders applications in time-critical diagnoses, such as strokes. Therefore, shortening the MRI scan time might help increase patient satisfaction, reducing motion artifacts from patient movement and saving the medical cost. In order to address the above issues, various acceleration techniques have been developed to speed up acquisition time under the premise of guaranteeing the image quality.

In recent years, inspired by deep learning (DL) [1], many techniques have been proposed to achieve efficient and accurate MRI reconstruction. Compared with traditional methods, the reconstruction quality of DL has been greatly improved, especially under very high acceleration rates [2-5]. DL methods mainly can be categorized as follows: Supervised and unsupervised schemes [6]. So far, most physics-based DL-MRI reconstruction approaches fall into the category of supervised learning. In general, one major challenge of supervised learning is the requirement of abundant high-quality reference images for network training. However, in practical settings, it may not be possible to acquire fully-sampled data as reference images due to the motion and imaging speed restriction. As a result, the requirement of reference images for network training may greatly restrict the broad applications of supervised learning in MRI reconstruction. To circumvent the above restriction, unsupervised learning is introduced for MRI reconstruction. Since generative models estimate the potential data distribution, it is widely recognized as a category of unsupervised deep learning.

More precisely, generative models can be divided into two main categories: One type that composing of the cost function-based models, such as variational autoencoders (VAEs) [7], auto-regressive models like PixelCNN [8], flow-based generative models like GLOW [9] and generative adversarial networks (GANs) [10]. The other category is the energy-based model (EBM) [11] where the joint probability is defined using an energy function. Particularly, the main goal of the VAEs algorithm is to approximate the data distribution using a latent variable model and then optimizes its parameters for a given set of examples using variational approximation. So far, VAE-based variants have been applied to MRI reconstruction. For example, Tezcan *et al*. [12] used VAEs as a tool for describing the data density prior (DDP). In the follow-up development, many researchers continued to carry out further researches on the basis. Liu *et al.* [13] employed enhanced denoising autoencoders (DAEs) as prior to MRI reconstruction (EDAEPRec).

Semi-amortized VAEs [14] use an encoder network followed by stochastic gradient descent on latent to improve the ELBO.

Autoregressive generative models like PixelCNN have also been verified in MRI reconstruction [15]. To the best of our knowledge, the PixelCNN, introduced by Oord *et al.* [8] is a powerful generative model of images with tractable likelihood, as the functional form is very flexible. More importantly, PixelCNN generally gives better performance in MRI reconstruction due to the fast-training speed. In addition, PixelCNN++ [16] is another architectural improvement of PixelCNN, with several beneficial modifications supported by experimental evidence. Besides of the above generative models, flow-based generative models [17-19] are also efficient to parallelize for both inference and synthesis.

Another promising direction to address unsupervised MRI reconstruction is using GANs that consisting of a discriminator and a generator. Goodfellow *et al*. [10] first proposed the framework of GAN and it have been proven very useful in constructing a generative model for unsupervised MRI reconstruction. For example, conditional generative adversarial network-based approach (DAGAN) [20] not only designed a U-Net network to reduce aliasing artifacts, but also coupled the adversarial loss with an innovative content loss. More recently, Bora *et al*. [21] proposed a framework, called Ambient-GAN, for learning generative models from underdetermined linear systems. They demonstrated encouraging results for small-scale simulated datasets. However, the generator network has no access to the real data and only learns through the discriminator's feedback. Therefore, training of GANs [22] is conducted in an alternation fashion that between the discriminator and the generator.

It is worth noting that another generative model has set off an upsurge, namely EBMs [23]. EBMs capture dependencies by associating a scalar energy (a measure of compatibility) to each configuration of the variables, and they provide a way to define an unnormalized probability density of images by a well-designed and trainable potential energy function. Moreover, the core idea of the model is to use the energy function to capture certain statistical properties of the input data and map it to energy. Fig. 1 vividly shows the frame structure of the EBM. Most importantly, the idea of combining energy-based models, deep neural network, and Langevin dynamics [24] provides an elegant, efficient, and powerful way to synthesize data with high quality. Accordingly, it establishes a new type of generative model framework, i.e., energy-based generative model, which is a new way to train deep neural networks in an unsupervised manner. Despite these benefits, the partition function of EBMs is generally intractable to calculate exactly. Moreover, although interest about EBMs is increasing thanks to several tricks aiding better scaling, its adaptation in MRI reconstruction has not been discussed, which leaves room for exploration.

In this work, one of the main purposes is to focus on the application of EBMs to MRI reconstruction, on the basic tool of Du *et al*. [25]. Among various advantages of EBMs, self-confrontation characteristic of the energy model is utilized in our work. Equipped with Langevin dynamics, it is comparable to or even better than other generative models on the reconstruction performance. On one hand, it does not require special model architecture, unlike auto-regressive and flow-based models. On the other hand, the maximum likelihood learning of the model does not suffer from issues like mode collapse in GAN or posterior collapse in VAE. Most notably, EBMs were only used for real-valued natural images in the past, while we exploit them to the complex-valued MRI reconstruction for the first time.

## Theory

In the previous section, we briefly introduce background on generative models, especially EBM. Motivated by the analysis mentioned above, striving to develop EBM to MRI reconstruction is a decent choice. EBM is an undirected model which enables itself to better approximate the data density than other generative networks. Simultaneously, EBM highlights some unique thoughts of self-adversarial learning which view the energy function as discriminator or generator that attributes energy to the data manifold regions. Nevertheless, there exist two major different characteristics between EBM and GAN in essence. In this section, we learn an energy-based prior for MR images and integrate it in the reconstruction model. Then we present a simplified special case of this generative model and compare it with GAN, detailed theoretical study will be presented in the following.

### Prior-based reconstruction in MRI

k-Space of MRI is related to image data by Fourier transform [26]. It is generated from spatial domain to Fourier domain by applying fast Fourier transform. In k-Space, central region consists of low frequency coefficients and peripheral region consists of high frequency coefficients. Before explaining the method in details, we will formalize the MRI process. Concretely, the partial k-Space raw MRI data $y$ can be obtained by the following equation:

$$y = F_P x + n \tag{1}$$

where $F_P = PF$ is a under-sampled measurement matrix, $F$ denotes the Fourier transform, $P$ is the sampling mask, $x$ is an MR image to be reconstructed, and $n$ is the measurement noise.

Since the sub-Nyquist sampling in the forward model Eq. (1) usually leads to ill-condition, the regularizer-based prior information is incorporated into the objective function for desirable reconstruction. The image recovery is needed to be formulated as an optimization problem:

$$\underset{x}{Min}\left\|F_P x-y\right\|^2+\lambda\log p(x) \tag{2}$$

where the first term enforces consistency between the acquired and reconstructed data, $\lambda$ is a regularization parameter and $\log p(x)$ enforces prior information to improve reconstruction performance.

## Prior learning with maximum likelihood estimation

Following the methodology in Eq. (2), we introduce EBM for learning the prior knowledge $\log p(x)$, and describe the maximum likelihood training scheme.

Given an input $x$ and the data distribution $P_D(x)$, the main goal of EBM is to learn an energy function $E_\theta(x)$ that assigns low energy values to inputs $x$ in the data distribution and high energy values to other inputs. In our work, this energy function is represented by a deep neural network parameterized with weights $\theta$. Furthermore, the energy function defines a probability distribution via the Boltzmann distribution:

$$p_\theta(x)=\exp(E_\theta(x))/Z(\theta) \tag{3}$$

where $Z(\theta)=\int\exp(-E_\theta(x))dx$ is the input-dependent normalizing partition function. For notational convenience, we treat the sample average and the expectation as identical. Hence, the log-likelihood is $\mathcal{L}_1=\mathbb{E}_{P_D(x)}[\log p_\theta(x)]$. In order to impel the Boltzmann distribution defined by energy function $E_\theta(x)$ to model the data distribution $P_D(x)$, it has been successfully trained with maximum likelihood estimation. Thus, the derivative of the log-likelihood is:

$$\mathcal{L}_1=\mathbb{E}_{P_D(x)}[E_\theta(x)-\log Z(\theta)] \tag{4}$$

**Lemma 1** [27]. Given a $\theta$-parametrized energy-based distribution $p_\theta(x)\propto\exp(-E_\theta(x))$ as well as supposing both $\exp(-E_\theta(x))$ and its partial derivative $\nabla_\theta\exp(-E_\theta(x))$ are continuous *w.r.t.* $\theta$ and $x$, we have the formation under *Leibniz integral rule*, $\forall x\in\mathbb{X}$:

$$\nabla_\theta\log p_\theta(x)=-\nabla_\theta E_\theta(x)+\mathbb{E}_{p_\theta(x)}[\nabla_\theta E_\theta(x)] \tag{5}$$

Because of the intractable partition function, we do not have access to the normalized density $p_\theta(x)$ and cannot directly minimize the objective of the maximum likelihood estimation. Fortunately, based on **Lemma 1** for derivation, we can further estimate the gradient of this objective:

$$\nabla_\theta\mathcal{L}_1=\mathbb{E}_{x^+\sim P_D}[\nabla_\theta E_\theta(x^+)]-\mathbb{E}_{x^-\sim p_\theta}[\nabla_\theta E_\theta(x^-)] \tag{6}$$

Intuitively, as shown in the left profile of Fig. 2, there is an interesting phenomenon that energy values of data samples are pushed down, while samples from the energy distribution are pushed up. More formally, the gradient decreases energy of the positive data samples $x^+$, while increasing the energy of the negative samples $x^-$ from the model $p_\theta(x)$. The detailed derivation of Eq. (6) is provided in **Appendix A**.

## Reconstruction with Langevin dynamics

Evaluating a universal energy function is convenient to the comparison of the relative probability for different inputs. Nevertheless, for most choices of the energy function $E_\theta(x)$, one cannot compute and even rely on the estimation of the partition function. As a result, estimating the normalized densities is intractable and estimating standard maximum likelihood of the parameters $\theta$ is not straightforward. One commonly used strategy is to use Markov Chain Monte Carlo (MCMC) [28] sampling to directly estimate the partition function, in which one iteratively updates a candidate configuration, until these configurations converge in distribution to the desired distribution. There are some well-established approximate methods based on MCMC such as random walk, Gibbs sampler [29], which have long mixing times especially if the energy function is complicated.

To improve the mixing time of the sampling procedure, we exploit MCMC with a Langevin dynamics so that the distribution $q_\theta(x)$ will approach to the model distribution $p_\theta(x)$, and this procedure generates samples from the distribution defined by the energy function. Additionally, model parameters are updated based on the maximum likelihood estimation, i.e.,

$$\nabla_\theta \mathcal{L}_{ML} = \mathbb{E}_{x^+ \sim P_D}[\nabla_\theta E_\theta(x^+)] - \mathbb{E}_{x^- \sim q_\theta}[\nabla_\theta E_\theta(x^-)] \tag{7}$$

As a gradient-based MCMC method, Langevin dynamics defines an efficient iterative sampling process, which asymptotically produces samples from an energy-based distribution:

$$\tilde{x}^t = \tilde{x}^{t-1} - \frac{\lambda}{2}\nabla_x E_\theta(\tilde{x}^{t-1}) + \omega^t, \omega^t \sim N(0,\lambda) \tag{8}$$

where the iterative procedure Eq. (8) defines a distribution $q_\theta(x)$ such that $\tilde{x}^T \sim q_\theta(x)$. As stated by Welling *et al.* [30], the distribution $q_\theta(x)$ of $\tilde{x}^T$ converges to the model distribution $p_\theta(x) \propto \exp(-E_\theta(x))$ when $\lambda \to 0$ and $T \to \infty$. Thus, samples are generated implicitly by the energy function $E_\theta(x)$ as opposed to being explicitly generated by a feedforward network. In summary, the procedure of sample generation can preclude explicit generator models limitations in a certain way.

Note that even few bounds on the mixing time are known, in practice, researchers found that using a relatively small $\lambda$ and finite $T$ suffices to produce good samples, and so far, various scalable techniques have been developed for the purpose of sampling from an EBM efficiently. For instance, Du *et al.* [26] proposed to use Langevin dynamics with a sample replay buffer to reduce mixing time and improve sample diversity, which is a sampling strategy we employ in the ex-periments. In details, we use a sample storage space $\mathbb{S}$ in which we preserve past generated samples $\tilde{x}$ and use either these samples or uniform noise to initialize Langevin dynamics procedure. It has the benefit of continuing to refine past samples, further increasing number of sampling steps $T$ as well as sample diversity.

More critically, arbitrary energy models can have sharp changes in gradients that can make sampling with Langevin dynamics unstable. Hence, two strategies are introduced to further improve the algorithm efficiency. One of the strategies is constraining the Lipschitz constant of the energy network by adding spectral normalization to all layers which can ameliorate these issues. The other key component is diminishing $L_2$ regularize energy magnitudes for both positive and negative samples during training. Thereby, it is possible to preserve the difference between positive and negative samples while avoiding actual values fluctuation to unstable value. Subsequently, the optimization objective can be modified as:

$$\Delta\theta \leftarrow \nabla_\theta \frac{1}{N}\sum\nolimits_n \beta(E_\theta(x_n^+)^2 + E_\theta(x_n^-)^2) + E_\theta(x_n^+) - E_\theta(x_n^-) \tag{9}$$

where parameter $\beta$ is the $L_2$ coefficient which can bound the magnitude of the unnormalized distribution.

By assembling the discussions in the above subsections, the final formulation of EBMRec model at each iteration of the annealed Langevin dynamics can be formulated as follows:

$$\underset{x}{Min}\left\|F_P x - y\right\|^2 + \lambda\left\|x - \tilde{x}^t\right\|^2 \tag{10}$$

Similar to the conventional iterative methods, we update the solution in a two-step alternative manner, via tackling the data-fidelity term and the regularization term subsequently. Thus, the least-square (LS) minimization in Eq. (10) can be solved as follows:

$$(F_p^T F_p + \lambda)x^t = F_p^T y + \lambda\tilde{x}^t \tag{11}$$

Let $F \in C^{M\times N}$ denotes the full Fourier encoding matrix which is normalized as $F^T F = 1^M$ . $Fx(k_v)$ stands for the updated value at under-sampled k-Space location $k_v$, and $\Omega$ represents the sampled subset of data, it yields,

$$Fx(k_v) = \begin{cases} F\tilde{x}^t(k_v), & k_v \notin \Omega \\ \dfrac{FF_p^T y(k_v) + \lambda F\tilde{x}^t(k_v)}{(1+\lambda)}, & k_v \in \Omega \end{cases} \tag{12}$$

To summarize, we conduct a comprehensive analysis and description of the algorithm used for MRI reconstruction based on EBM. Strictly speaking, the maximum likelihood learning of EBM is essentially the “analysis by synthesis” scheme that called by Grenander *et al.* [31]. Within each learning iteration, we generate synthesized examples by sampling from the current model, and then update the model parameters based on the difference between the synthesized examples and the observed examples, so that eventually the synthesized examples match the observed examples in terms of some statistical properties defined by the model. The detailed description of EBMRec procedure (**Algorithm 1**) is provided in **Appendix B.**

## Re-visiting EBM from adversarial perspective

As pointed out in previous chapter, EBM uses the energy function to contrast true data and samples generated by $E(x)$ itself implicitly through MCMC, while the most popular generative model GAN uses the discriminator to contrast true data and samples generated by the generator. Therefore, the training procedure of EBM is essentially an implicit self-adversarial game.

More precisely, GAN with minimax game has achieved state-of-the-art performance in many image generation tasks. The basic idea of GAN is to simultaneously train a generator $G$ and a discriminator $D$. The configuration of GAN is illustrated in the right profile of Fig. 2. The input of the generator is a pure random noise sampled from a prior distribution. Accordingly, the output is expected to have visual similarity with the real sample that is drawn from the real data distribution. As for the discriminator, the input of it is either a real or generated sample while the output is a single value indicating the probability of the input being a real or fake sample. From the game theory point of view, the training objectives can be expressed mathematically as:

$$\underset{G}{Min}\max_{D} V(D,G) = \mathbb{E}_{x\sim P_D(x)}[\log D(x)] + \mathbb{E}_{z\sim p_z(z)}[\log(1-D(G(z)))] \tag{13}$$

Ideally, GAN has shown outstanding performance in image-to-image translation and has been employed for reconstructing under-sampled MRI. Nonetheless, a problem commonly faced in training GANs is mode collapse when the distribution learned by generator focuses on a few limited modes of the data distribution. Hence, its research is still in its early stage and has not been fully exploited in MRI reconstruction. Especially, some recent GAN variants of GANs including RefineGAN [32], WGAN [33], BiGAN [34] and so on mainly benefit from the discriminator ability. In general, they are trained in an end-to-end manner and not indeed unsupervised methods.

The essence of the EBM is to build an energy function that maps each point of an input space to a single scalar. It follows the underlying idea of generative adversarial game, spiritually similar to GAN. As shown in the right profile of Fig. 2, since an energy model is the only designed and trained object, it requires fewer model parameters than GAN that uses two networks. More importantly, the sample generation is an iterative stochastic optimization process, which allows for a trade-off between generation quality and computation time. Although it is visually described from the right profile of Fig. 2 that both EBM and GAN seem to follow the idea of adversarial game, there are substantial differences between them upon closer consideration, which can be summarized as follows:

**i.** Rather than GAN that employs the minimax game to comply with alternative training between $\nabla_\theta G$ and $\nabla_\theta D$, EBM only trains an energy model $\nabla_\theta E$ and then minimaximizes the energy of positive data samples $E(x^+)$ and negative samples $E(x^-)$ at the same time. It thus indicates that training EBM is an effectively self-adversarial procedure, rather than the adversarial strategy that against each other in GAN. Particularly, at the adversarial training, samples are generated implicitly by the Langevin dynamics that involves the energy gradient $\nabla_x E$. Hence, it mirrors a typical image generator network architecture but without the necessary to be explicitly designed or balanced such as GAN. In summary, the training objective of GAN is $\{\nabla_\theta G, \nabla_\theta D\}$, while EBM is $\{\nabla_x E, \nabla_\theta E\}$.

**ii.** GAN uses feed-forward generative network to obtain samples in an end-to-end manner, while EBM uses the Langevin dynamics to generate samples in an iterative scheme. Crucially, with an implicit iteration generation and no gradient through sampling are important as it controls the diversity in likelihood models and the mode collapse in GAN.

## Methods

### MRI data and simulation settings

We conduct experiments on five commonly used image datasets to demonstrate the proposed algorithm, i.e., brain images from *SIAT*, knee images from *FastMRI*, and three multi-coil acquisition data *Test1* of MoDL [35], *Test2* of DDP [36] and in vivo dataset of LINDBERG [37].

Firstly, the brain images are selected from *SIAT* dataset, which is provided by Shenzhen Institutes of Advanced Technology, the Chinese Academy of Science. Informed consents are obtained from the imaging subject in compliance with the institutional review board policy. The collected dataset includes 500 2D complex-valued MR images with size of $256 \times 256$. In details, MR images are acquired from a healthy volunteer by a 3.0T Siemens Trio Tim MRI scanner using the T2 weighted turbo spin echo sequence (repetition time (TR)/echo time (TE)=6100/99 $ms$, $220 \times 220$ $mm^2$ field of view). Besides, the voxel size is $0.9 \times 0.9 \times 0.9$ $mm$.

Secondly, experiments are also conducted on single coil knee images from *FastMRI* dataset. The following sequence parameters are used: Echo train length 4, matrix size $320 \times 320$, in-plane resolution $0.5\ mm \times 0.5\ mm$, slice thickness $3\,mm$, with no gap between slices. Timing varies between systems, TR ranging between 2200 and 3000 milliseconds, and (TE) between 27 and 34 milliseconds. We randomly select 80 files (about 2941 images) and 2 files (about 50 images) from *FastMRI* to our model, respectively. Notably, the training set is generated in the same way as *SIAT* dataset, which produces 26469 patches via affine transformation.

Finally, besides of the above single coil datasets, experiments are conducted on multi-coil acquisition data *Test1* of MoDL and *Test2* of DDP to verify the performance of EBMRec for parallel imaging reconstruction. Additionally, in order to further explore the advantage of EBMRec in robustly handling the sensitivity information for calibration-free parallel imaging reconstruction, the ablation experiment is also conducted on another multi-coil acquisition data in vivo dataset of LINDBERG. Quality of the MR image reconstructed from partial k-Space is measured in terms of the popular PSNR (Peak Signal to Noise Ratio, dB) and the powerful perceptual quality metrics SSIM (Structural Similarity). Especially, PSNR describes the relationship of the maximum possible power of a signal with the power of noise corruption. SSIM measures the similarity between the original image and reconstructed images on three aspects: Luminance, contrast, and structural correlation.

## Network architecture and model parameters

EBM typically aims to learn an energy function that produces the best dependencies from the input variables to a scalar energy. Recently, by using neural network as the energy function, many approaches under EBM are able to model complex data such as images and texts [38-39]. Considering that residual network has become one of the most successful neural networks in image processing community, it is used as an energy function in this work. Detailed architecture flowchart and concrete parameter settings in residual model are illustrated in Supporting Information Fig. S1.

As shown in Fig. S1, to achieve good non-linearity approximation, Conv layer and ResBlock are adopted in residual model. In the ResBlock, feedforward neural networks with "shortcut connections", which means one or more layers are skipped. More precisely, the shortcut connections simply perform identity mapping, and their outputs are added to the outputs of the stacked layers. Intuitively, one can observe that the filter number in Conv layer of the model is 64 with a kernel size of $3\times3$. However, the model under the ResBlock down layer will change the number of filters and increase regularly as the network deepens. Last but not least, the number of filters in the last Global Sum Pooling layer and Dense layer are both 1.

During the learning phase, we use fully sampled MR images as the network input and disturb it simultaneously via random Gaussian noise of various amplitudes. For *SIAT* and *FastMRI* datasets, it is worth noting that the input and output are all complex-valued images with the same size and each includes real and imaginary components. Adam is selected as an optimizer with the learning rate 0.0003. Subsequently, EBMRec model is trained after 180000 iterations with the batch size of 16 that takes around 72 hours. It is performed with Pytorch interface on 2 NVIDIA Titan XP GPUs, 12GB RAM. For the convenience of reproducibility, source code is available at: https://github.com/yqx7150/EBMRec.

# Results

## Compressed sensing reconstruction

To evaluate the performance of EBMRec in the case of single coil reconstruction, 31 complex-valued brain images from *SIAT* and 50 complex-valued knee images from *FastMRI* are used in this subsection. Moreover, several state-of-the-art methods are compared with EBMRec, including patch-based DLMRI [40], enhanced denoising autoencoder based EDAEPRec [13], and RefineGAN [32]. We additionally implement low-rank modeling of local k-Space neighborhoods LORAKS [41], supervised learning U-Net [42], and multi-channel enhanced deep mean-shift prior MEDMSPRec [43] on *FastMRI*. The quantitative comparisons among the other algorithms and EBMRec are recorded in Table 1.

Qualitatively, the average PSNR result of 31 test images obtained by EBMRec is 2 dB higher than the iterative algorithm DLMRI. Furthermore, EBMRec yields the highest values in the most majority of the sampling rates. For example, for 15% retained data, most of the aliasing artifacts have still been suppressed effectively and we can still obtain an average of PSNR over 30 dB, but there is obvious loss of structural details such as organ edges. The principal reason for this phenomenon is the significant information loss in the low frequency regions. In contrast, other methods are dramatically affected by higher accelerated factors. Before moving on, it is worth mentioning that with the increase of acceleration factor, the performance of EBMRec is much more dominant. In other words, under the extremely lack of data acquisition, EBMRec still reconstructs good results. Moreover, in terms of PSNR values, EBMRec achieves better quantitative results under 1D Cartesian samplings at $R=4$ on *FastMRI* dataset. Theoretically, the inferiority of LORAKS indicates that the traditional model is insufficient to tackle the challenging data with amounts of high-resolution features and details. In summary, the proposed method EBMRec outperforms LORAKS, U-Net and MEDMSPRec.

Besides of the quantitative comparison, the visual quality is also highlighted. As can be seen in Figs. 3-4, visual quality of reconstructions for different methods varies. The results of DLMRI are less recognizable because it tends to preserve the shape of the object while sacrificing the texture and structure. On the contrary, EBMRec reaches a balance between the shape and texture as shown in Fig. 3. Besides, error maps demonstrated in Figs. 3-4 indicate that EDAEPRec is better than RefineGAN method in terms of more abundant recovered edge details. However, it still suffers from some undesirable artifacts and loses details. LORAKS can improve the reconstruction per-formance compared to zero-filled image, but it is hard to see a significant improvement when an amount of aliasing artifact is presented. U-Net has excellent reconstruction effects in some areas, but the overall reconstruction quality is still worse than EBMRec. To sum up, EBMRec can achieve more satisfactory results with clearer contours, sharper edges, and finer image details under various sampling masks. For instance, the proposed method shows slightly better preservation of boundaries between gray matter and white matter. More importantly, the proposed method uniquely supports different under-sampling trajectories without retraining the deep learning model.

**Parallel imaging reconstruction**

As well known, flexibility and robustness are the main characteristics of unsupervised learning strategy that differ from the supervised learning counterpart. In this subsection, EBMRec is compared with MoDL and DDP for parallel imaging reconstruction, respectively. It is worth noting that the test dataset contains multi-coil brain MR images, while EBMRec model is still trained on the single coil *SIAT* dataset. Table 2 tabulates the comparison of MoDL, DDP and EBMRec. Among them, MoDL model is trained on 360 brain MR images (12 coils, complex-valued image). MRI data and Cartesian readouts are acquired using a 3D T2 CUBE sequence and a 12-channel head coil, respectively. The matrix dimensions are $256\times232\times208$ with $1\,mm$ isotropic resolution. The coil sensitivity maps are estimated from the central k-Space regions of each slice using ESPIRiT [44] and are assumed to be known during experiments. Thus, the data has dimensions in rows × columns × coils as $256\times232\times12$. For DDP, it is trained on 790 (single coil, no phase) central T1 weighted slices with $1\,mm$ in-slice resolution from the HCP dataset and tested on an image from a volunteer acquired for this study (15 coils, complex-valued image) along with the corresponding ESPIRiT coil maps.

From the metrics in Table 2, it can be observed that the average PSNR values of the reconstructed each image by using EBMRec are higher than every other model. Particularly, as the under-sampling rate increases, the performance gain becomes more striking. This phenomenon further indicates that EBMRec is more advantageous and effective in severely ill-posed circumstances. To further prove the superiority of EBMRec, visualization results reconstructed by different methods with different acceleration factors are provided in Figs. 5-6. In general, EBMRec achieves faithful reconstruction with clear textures and boundaries. For example, compared to MoDL and DDP reconstruction, the proposed model demonstrates superior reconstructed details that refraining from significant aliasing artifacts.

**Calibration-free parallel imaging reconstruction**

Calibration-free techniques have presented encouraging performances due to their capability in robustly handling the sensitivity information. To further validate the feasibility of the proposed method, we implement the calibration-free parallel imaging under 2D Poisson disk under-sampling scheme at different acceleration factors against the state-of-the-art method, namely the calibration-free method with joint-sparse codes LINDBERG. In details, comparison experiments are conducted on different in vivo dataset with size of $256 \times 256$, which is provided by reconstruction database [37]. For fair comparison to LINDBERG, we empirically tune the parameters in their suggested ranges to give their best performances. Moreover, the images are of a great diversity with different number of receiver coils such as 4, 8 and 12.

Table 3 depicts the quantitative measurement comparison for these methods with respect to different acceleration factors and different number of receiver coils. It lists the PSNR and SSIM values between the degraded image and the original image with different acceleration factors. We can observe that the PSNR values of EBMRec are obviously higher than LINDBERG among different acceleration factors and number of receiver coils. Even in the case of acceleration factor $R=6$, EBMRec still produces reasonable result. Additionally, the SSIM values of EBMRec are closer to 1 when the acceleration factor $R=3$ and number of receiver coils are 4 and 8, which indicates that EBMRec has good performance with the best structural similarity between reference image and result. Visual quality of reconstructions for different methods also varies. As can be seen from Fig. 7, under acceleration factor $R=4$, EBMRec successfully removes structural artifacts while LINDBERG fails to remove the artifacts. Indeed, EBMRec is slightly better than LINDBERG in terms of reconstruction error maps and metrics. Furthermore, LINDBERG has some artifact noise at the edges of the image from the error maps. Interestingly, the same conclusion can be reached from the enlarged number of receiver coils and reconstruction errors that EBMRec exhibits the least error.

## Discussions

The most important motivation for this study is to reconstruct highly under-sampled MRI k-Space data accurately, which can shorten the scanning time effectively. With the experimental results discussed above, we have demonstrated that the proposed method can reliably and consistently recover the nearly aliased-free images with relatively high acceleration factors. Furthermore, as expected, the reconstruction results from energy-based prior show the successful preservation of the detailed anatomical structures such as texture and edges.

Interestingly, the approach for solving Eq. (2) converges quickly. The convergence tendency of PSNR curve versus iteration for reconstructing the 5th brain image is plotted in Fig. 8. It can be seen that the curve is wavy at early iterations and then becomes stable. Additionally, the computational costs of EBMRec and the competing methods are evaluated in Supporting Information

Table S1. As a result, with GPU implementation, the computational cost of our algorithm is moderate.

How different initial values would affect the efficacy of the proposed method to reconstruct MR images are investigated. To this end, two different initializations are used respectively, namely initializing with the uniform noise $x \sim N(-1,1)$ and zero-filled data. Supporting Information Table S2 presents the results produced by EBMRec with two initializations. It is noticed that if the zero-filled image is used for initialization, the results gained by EBMRec are almost the same regardless of initializations. This evidence illustrates that EBMRec is insensitive to initialization.

## Conclusions

The central contribution of our work is to investigate how to use the strategy in EBM for MRI reconstruction. In short, based on maximum likelihood estimation and Langevin dynamics inference, a new method of image reconstruction from incomplete measurements using energy-based generative prior is proposed. Firstly, it provides new insights to understand the maximum likelihood estimation of EBM which trains the model through a self-adversarial mechanism. Secondly, a corresponding iterative approach is developed to strengthen EBM training with the gradient of energy network. Finally, instead of sampling the partition function through amortized generation, MCMC with a Langevin dynamics is used for more efficient generating. Comprehensive experiment results demonstrated that EBMRec achieved superior performance. More importantly, EBMRec is generalizable for most reconstruction scenarios. Another interesting avenue for future research is examining high-dimensional space strategy and more imaging modalities for CT and PET reconstruction.

## Acknowledgements

The authors would sincerely thank Dr. Yilun Du for the previous work on the deep energy-based models and codes [25] that are very helpful in this paper. This work was in part supported by National Natural Science Foundation of China (61871206, 61601450), Basic Research Program of Shenzhen (JCYJ20150831154213680), and the Natural Science Foundation of Jiangxi Province (20181BAB202003).

# Appendix A: Derivation for the self-adversarial learning in Equation 6

For a log-likelihood function that's already defined as:

$$\mathbb{E}_{x\sim P_D(x)}[\log p_\theta(x)]$$

It should be as small as possible and then the derivative of the log-likelihood is as follow:

$$L_\theta = \mathbb{E}_{x\sim P_D(x)}[-\log p_\theta(x)]$$

Therefore, we minimize the negative log-likelihood of the data by gradient descent optimization algorithm:

$$\begin{aligned}
\nabla_\theta \log P_\theta(x) &= \nabla_\theta \log e^{-E_\theta(x)} - \nabla_\theta \log Z_\theta \\
&= -\nabla_\theta E_\theta(x) - \frac{1}{Z_\theta}\nabla_\theta Z_\theta \\
&= -\nabla_\theta E_\theta(x) - \frac{1}{Z_\theta}\nabla_\theta \int e^{-E_\theta(x)} dx \\
&= -\nabla_\theta E_\theta(x) + \frac{1}{Z_\theta}\int e^{-E_\theta(x)} \nabla_\theta E_\theta(x) dx \\
&= -\nabla_\theta E_\theta(x) + \int \frac{e^{-E_\theta(x)}}{Z_\theta} \nabla_\theta E_\theta(x) dx \\
&= -\nabla_\theta E_\theta(x) + \mathbb{E}_{x\sim P_\theta(x)}[\nabla_\theta E_\theta(x)]
\end{aligned}$$

Then, we have

$$\nabla_\theta L_\theta = \mathbb{E}_{x\sim P_D(x)}[\nabla_\theta E_\theta(x)] - \mathbb{E}_{x\sim P_\theta(x)}[\nabla_\theta E_\theta(x)]$$

Similarly, it means the optimize objective rewritten as:

$$\theta \leftarrow \theta - \varepsilon(\mathbb{E}_{x\sim P_D(x)}[\nabla_\theta E_\theta(x)] - \mathbb{E}_{x\sim P_\theta(x)}[\nabla_\theta E_\theta(x)])$$

## Appendix B: MRI reconstruction with energy-based prior model

**Algorithm 1** Reconstruction algorithm with energy-based prior

**Input**:

$P_D(x)$ - dataset, $x_n^+ \sim P_D(x)$

$x_n^-$ - $\mathbb{S}$ with 95% probability and uniform noise, $\mathbb{S} \leftarrow \varnothing$

$y$ - k-Space data

**Output**:

$E_\theta(x)$ - trained EBM

$x$ - the restored image

Give a random initial value $x^0$, $\sigma \in \{\sigma_i\}_{i=1}^{I}$, $\varepsilon$, $a$, $T$ ▷ Initialization

**while** $i \leftarrow 1$ to $I$ **do** ▷ Iteration

  Optimize objective $\beta\mathcal{L}_1 + \mathcal{L}_{ML} : \Delta\theta \leftarrow \nabla_\theta \frac{1}{N}\sum_n \beta(E_\theta(x_n^+)^2 + E_\theta(x_n^-)^2) + E_\theta(x_n^+) - E_\theta(x_n^-)$

  Get $\Delta\theta$ using Adam optimizer

  Pick a step size $\alpha_i = \varepsilon \cdot \sigma_i^2 / \sigma_I^2$

  **while** sample step $t \leftarrow 1$ to $T$ **do**

    Split $x$ into pieces for feeding to network

    Update $\tilde{x}^t \leftarrow \tilde{x}^{t-1} - \nabla_x E_\theta(\tilde{x}^{t-1}, \sigma_i) + \varpi, \quad \varpi \sim N(0,\sigma)$

    Merge $\mathbb{S} \cup x_n^-$ and $x_n^-$ into $\mathbb{S}$ and $x^0$ for projection

    Least-square reconstruction to obtain $x^t$

**Return** $x_n^-$

# Captions

**FIG. 1.** Overview of the MRI reconstruction. (a) In this method, we use the concept of energy-based model which maps different kinds of inputs into scalar energy via neural network. Meanwhile, parameter $\theta$ obtained by maximum likelihood estimation and self-adversarial cogitation. (b) After updating image, the generative image with data-fidelity term is projected onto $\{x \mid y = F_P x + n\}$ in Eq. (1).

**FIG. 2.** Detailed comparison of characteristics and structures in the flow chart of GAN and EBM. While both GAN and EBM follow the idea of self-confrontation, EBM only needs to optimize a model itself, thus avoiding the defects of model collapse. The right region depicts the change of energy curve during the training procedure in the EBM.

**FIG. 3.** Complex-valued reconstruction comparison of different methods on *SIAT* dataset, under pseudo radial sampling at acceleration factor $R=5$ in single coil and 256 × 256 matrix size. The intensity of error maps is five times magnified. The proposed method EBMRec substantially reduces the aliasing artifact and preserves image details.

**FIG. 4.** Complex-valued reconstruction results on *FastMRI* dataset at 1D Cartesian under-sampling pattern in single coil. From top to bottom: The reconstruction results on two different knee images at acceleration factor $R=4$. The intensity of error maps is five times magnified. From left to right: Ground truth, LORAKS, U-Net, MEDMSPRec, and EBMRec.

**FIG. 5.** Complex-valued reconstruction results on brain images at two 1D Cartesian under-sampling percentages in 15 coils parallel imaging. The intensity of error maps is five times magnified. From top to bottom: the reconstruction results on brain images at $R=2$ and $R=3$, respectively. From left to right: Ground truth, reconstruction by Zero-filled, DDP and EBMRec.

**FIG. 6.** Complex-valued reconstruction results on brain images at $R=6$ pseudo random sampling in 12 coils parallel imaging. The intensity of error maps is five times magnified. From left to right: Ground truth, reconstruction by Zero-filled, MoDL and EBMRec.

**FIG. 7.** Visual measurements of reconstructing in vivo dataset at $R=4$ 2D Poisson disk under-sampling pattern in calibration-free parallel imaging by LINDBERG and EBMRec. The intensity of error maps is five times magnified. From top to bottom: The reconstruction results in 8 coils and 12 coils, respectively.

**FIG. 8.** Quantitative PSNR curve versus iteration for reconstructing the 5th brain image. The convergence tendency reflects the stability of iterative steps.

**Table 1.** PSNR and SSIM comparison of different methods for single coil reconstruction under different under-sampling patterns at a variety of acceleration factors. Among them, 31 and 50 test images are used for single coil reconstruction including brain and knee MRI.

**Table 2.** Quantitative PSNR and SSIM measures for different algorithms with varying accelerate factors and sampling masks. Among them, DDP results of test images at $R=2$ and $R=3$ 1D Cartesian under-sampling pattern are obtained in 15 coils parallel imaging, while reconstruction MoDL results of 10 brain test images at $R=6$ 1D pseudo random sampling in 12 coils parallel imaging.

**Table 3.** PSNR and SSIM comparison of in vivo dataset for calibration-free parallel imaging at 2D Poisson disk under-sampling with varying accelerate factors and different number of receiver coils. From top to bottom: the proposed method EBMRec and LINDBERG, respectively.

## List of supporting information

**Supporting Information Figure S1.** The diagram shows the residual model with multiple residual network architectures and parameters, which is the prior model used in this study.

**Supporting Information Table S1.** Computational costs of EBMRec and the competing methods. Notice that the reference image is a $256\times256$ MRI image. DLMRI is executed with CPU, while RefineGAN, EDAEPRec and EBMRec are with GPU.

**Supporting Information Table S2.** Performance of reconstructing 31 test images by different initial values at radial sampling trajectories in single coil.

## Figures

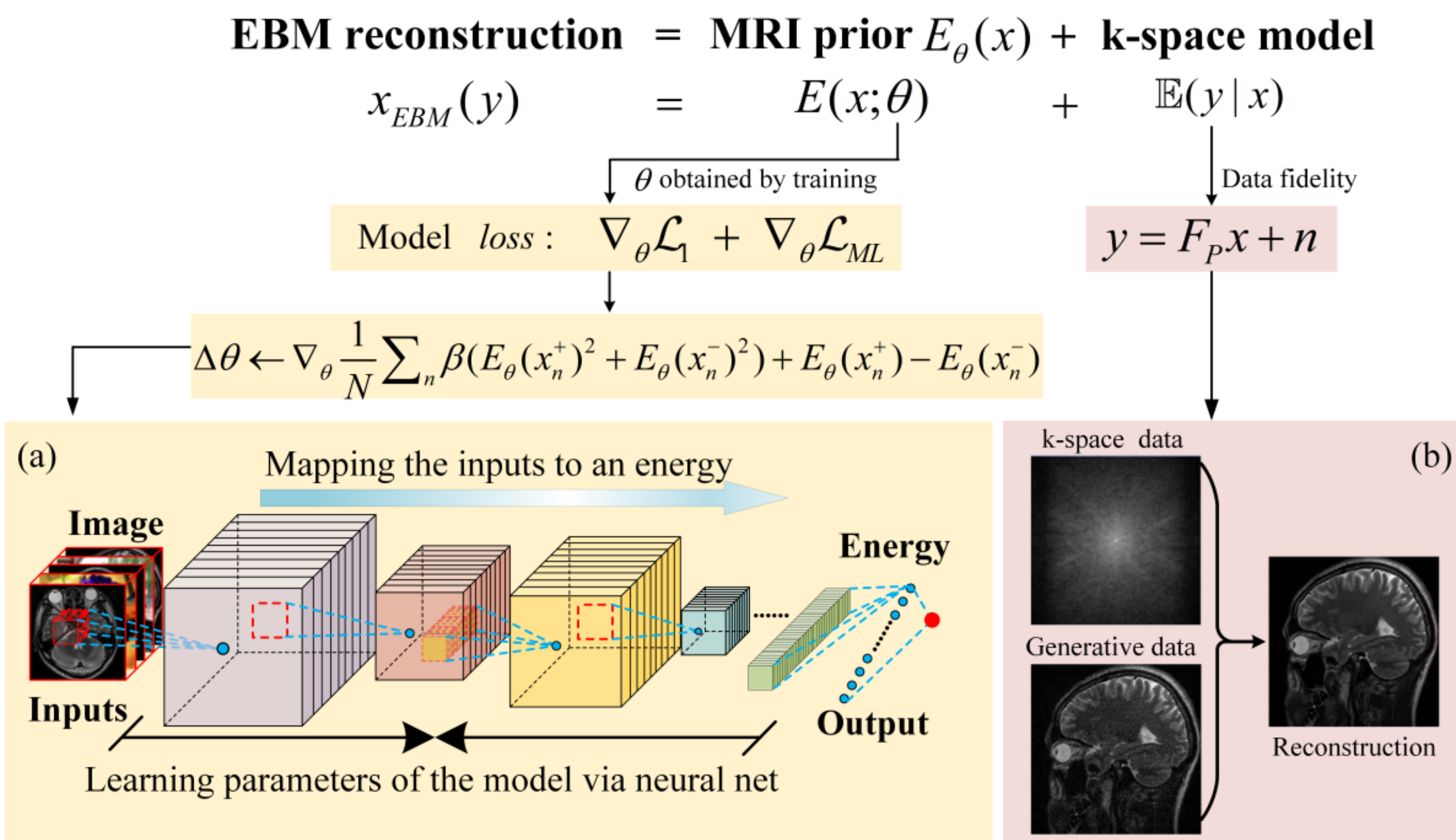

**FIG. 1.** Overview of the MRI reconstruction. (a) In this method, we use the concept of energy-based model which maps different kinds of inputs into scalar energy via neural network. Meanwhile, parameter $\theta$ obtained by maximum likelihood estimation and self-adversarial cogitation. (b) After updating image, the generative image with data-fidelity term is projected onto $\{x | y = F_P x + n\}$ in Eq. (1).

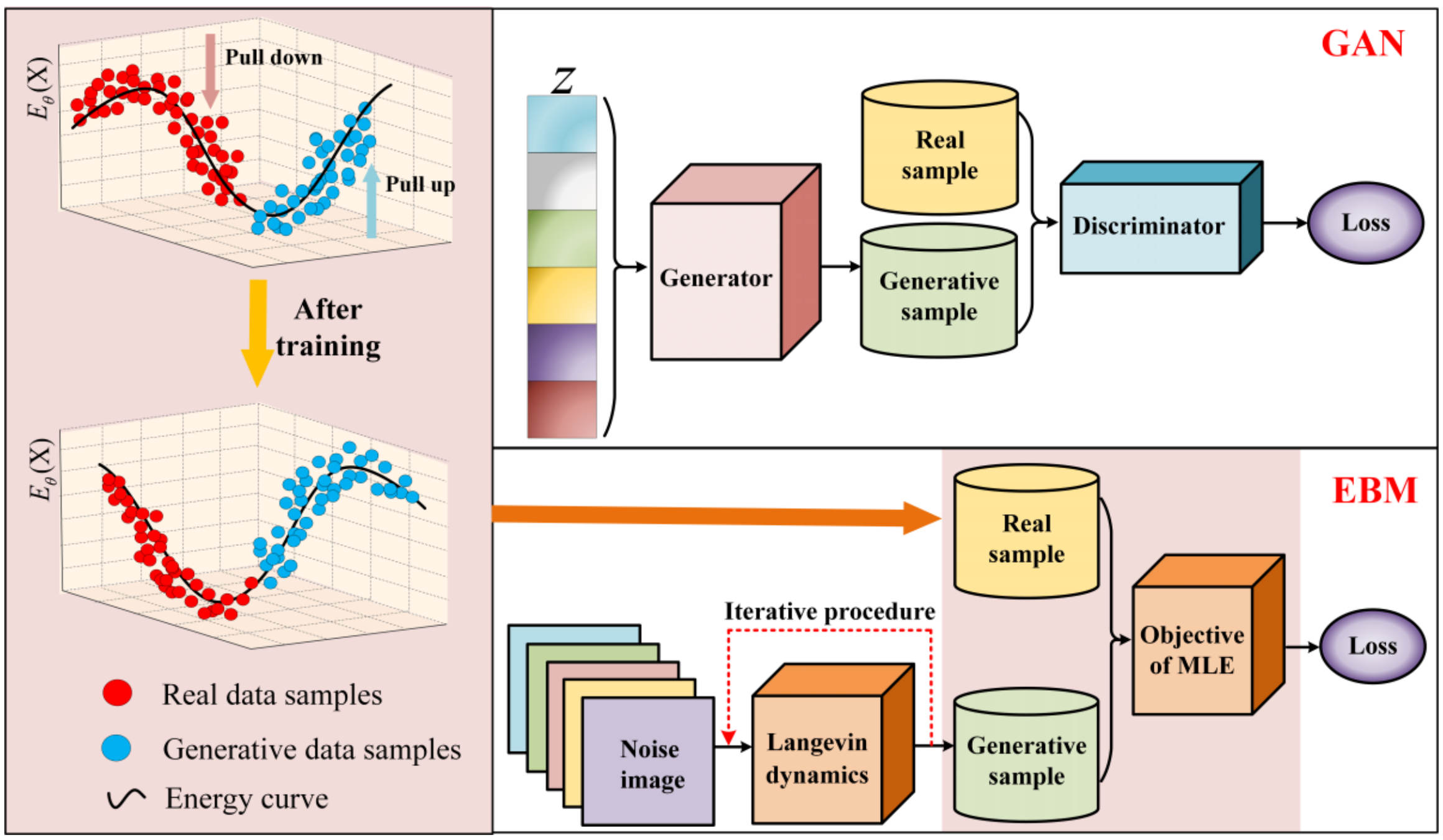

**FIG. 2.** Detailed comparison of characteristics and structures in the flow chart of GAN and EBM. While both GAN and EBM follow the idea of self-confrontation, EBM only needs to optimize a model itself, thus avoiding the defects of model collapse. The right region depicts the change of energy curve during the training procedure in the EBM.

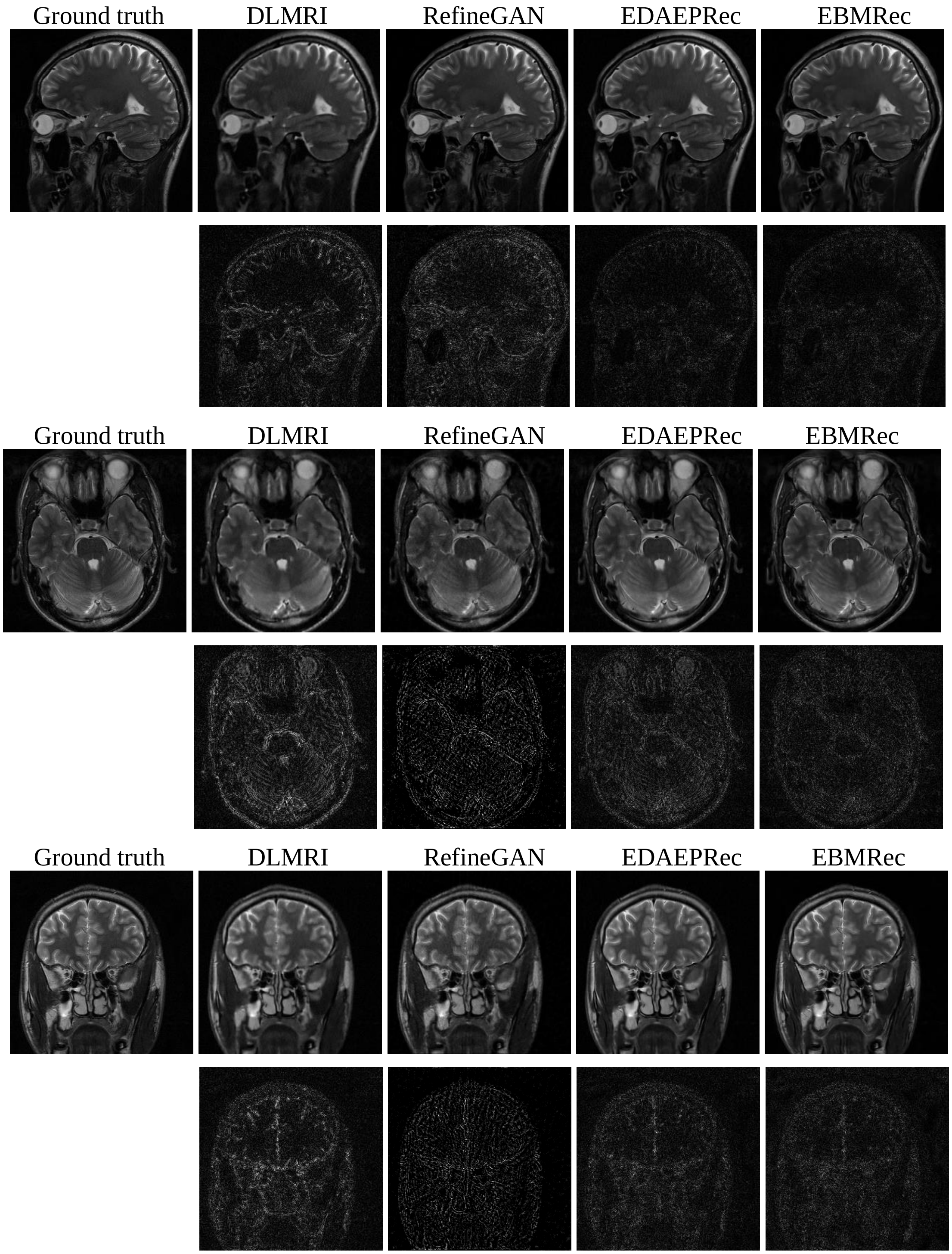

**FIG. 3.** Complex-valued reconstruction comparison of different methods on *SIAT* dataset, under pseudo radial sampling at acceleration factor $R=5$ in single coil and 256 × 256 matrix size. The intensity of error maps is five times magnified. The proposed method EBMRec substantially reduces the aliasing artifact and preserves image details.

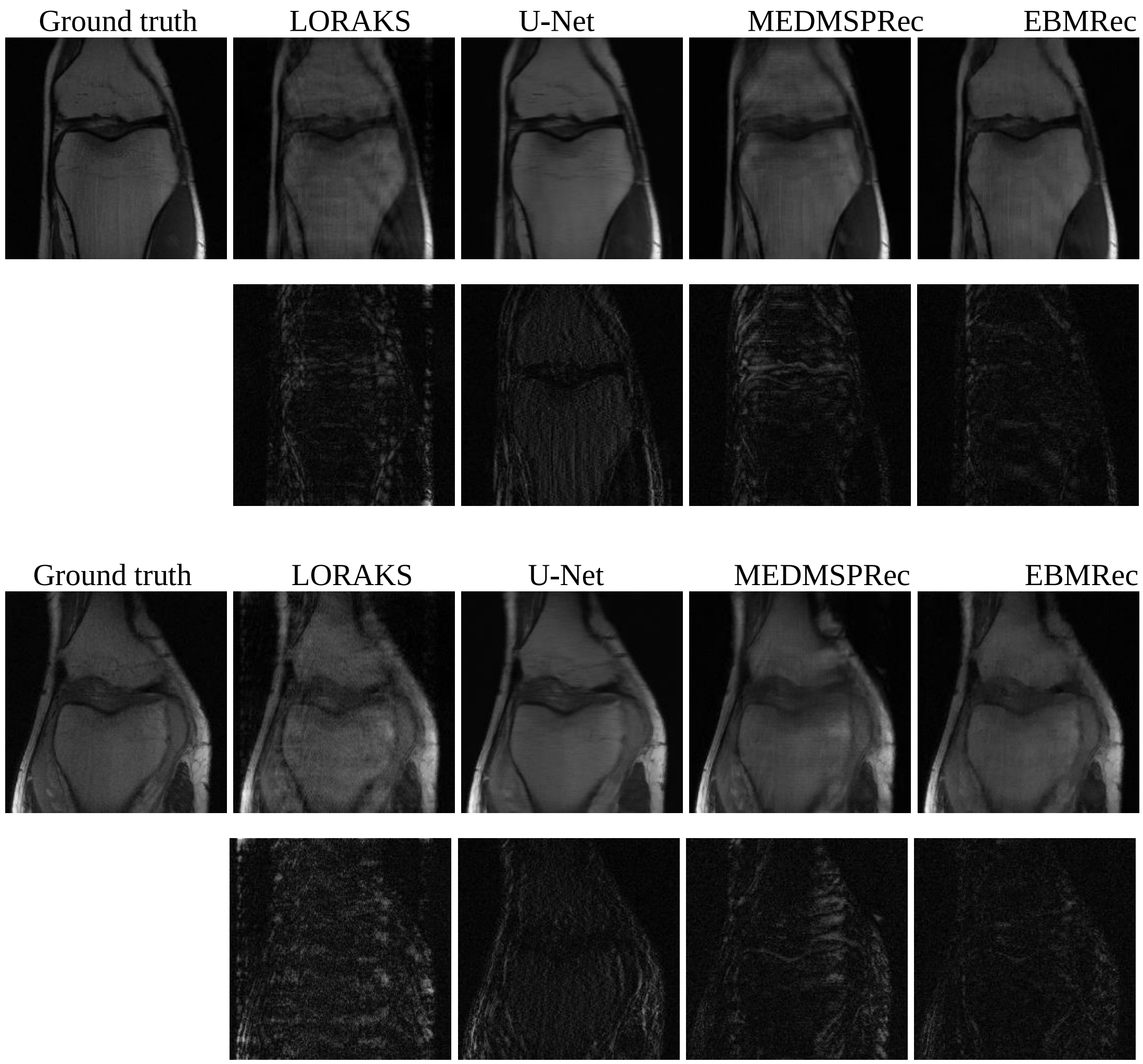

**FIG. 4.** Complex-valued reconstruction results on *FastMRI* dataset at 1D Cartesian under-sampling pattern in single coil. From top to bottom: The reconstruction results on two different knee images at acceleration factor $R = 4$. The intensity of error maps is three times magnified. From left to right: Ground truth, LORAKS, U-Net, MEDMSPRec, and EBMRec.

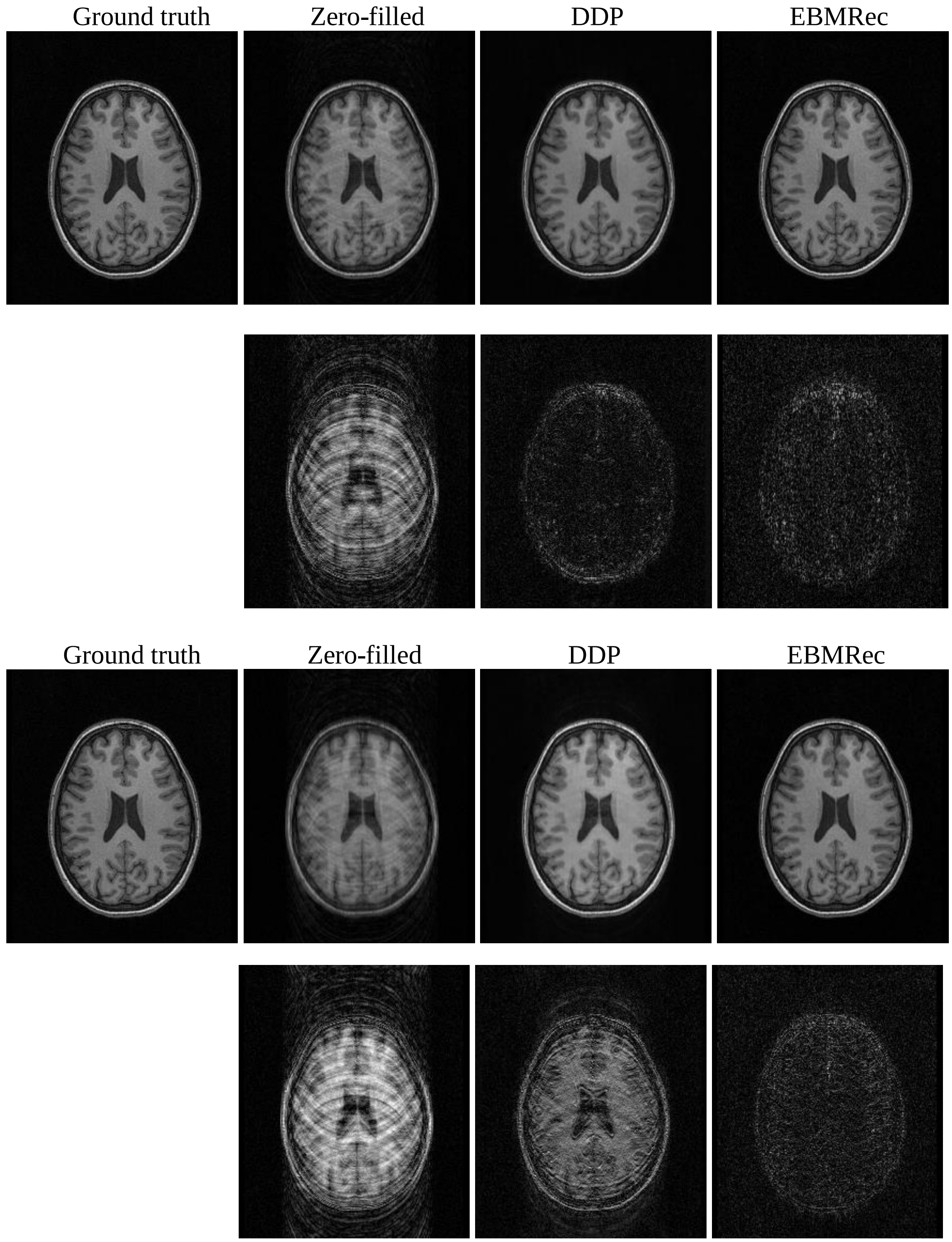

**FIG. 5.** Complex-valued reconstruction results on brain images at two 1D Cartesian under-sampling percentages in 15 coils parallel imaging. The intensity of error maps is five times magnified. From top to bottom: the reconstruction results on brain images at $R=2$ and $R=3$, respectively. From left to right: Ground truth, reconstruction by Zero-filled, DDP and EBMRec.

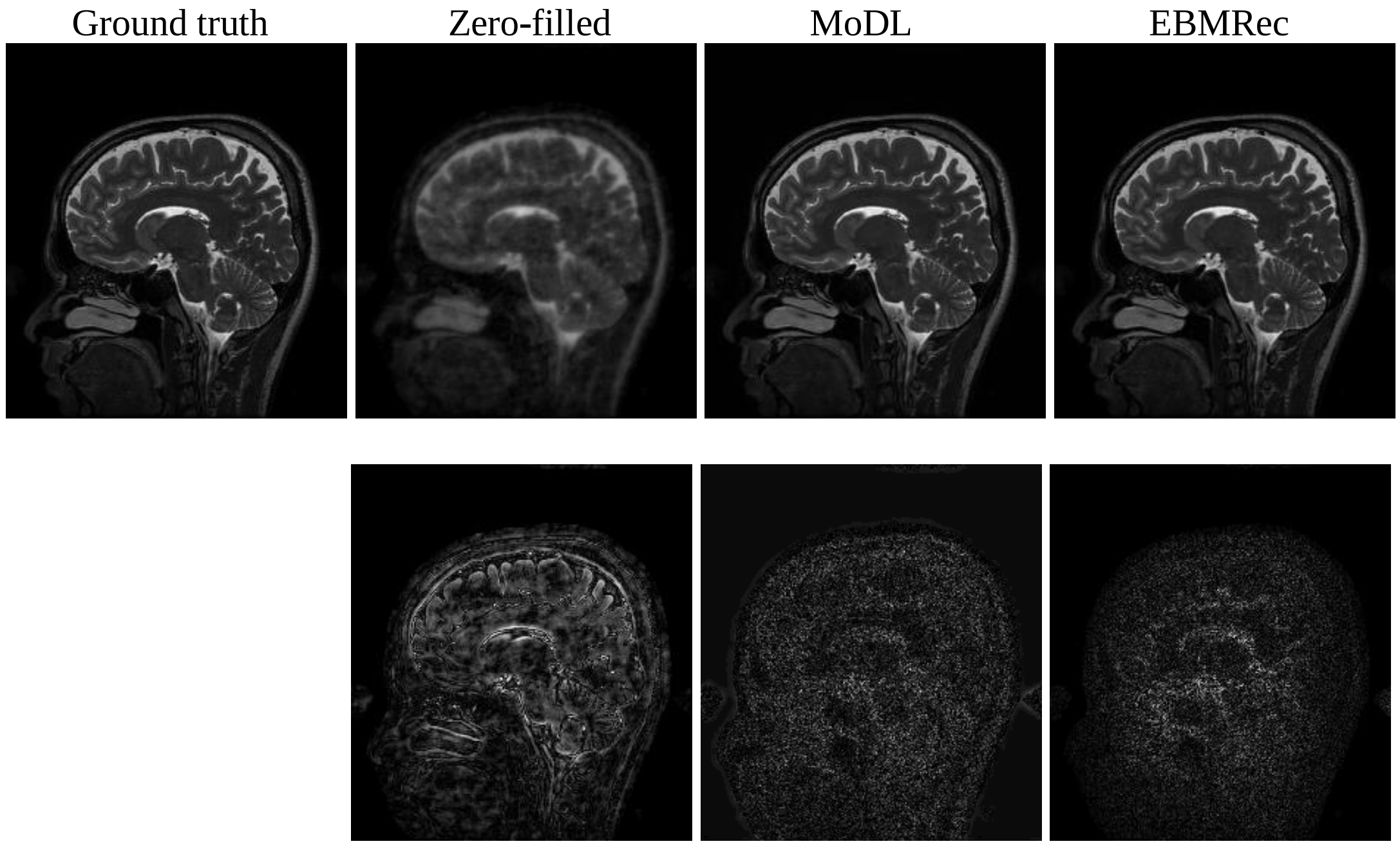

**FIG. 6.** Complex-valued reconstruction results on brain images at $R = 6$ pseudo random sampling in 12 coils parallel imaging. The intensity of error maps is five times magnified. From left to right: Ground truth, reconstruction by Zero-filled, MoDL and EBMRec.

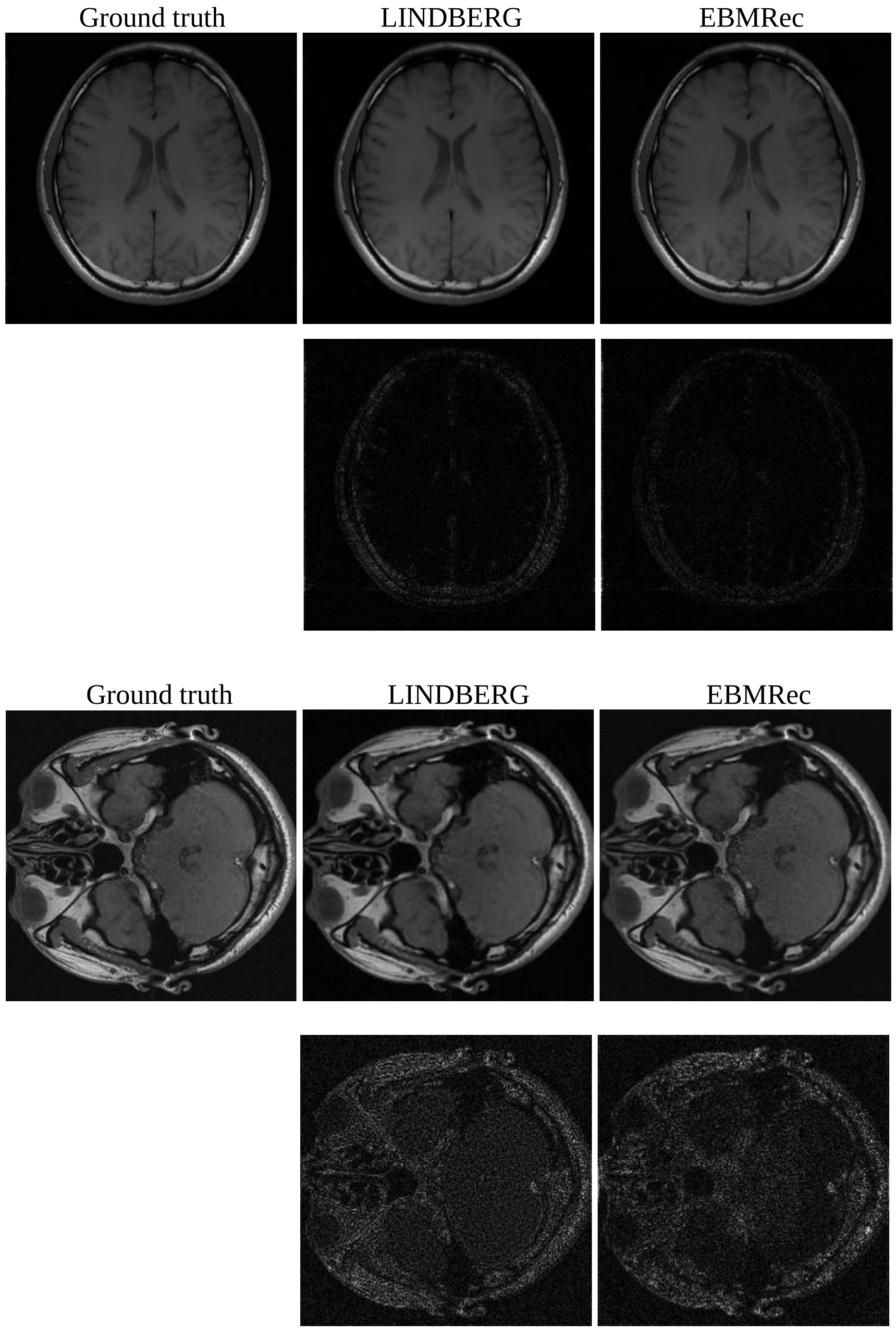

**FIG. 7.** Visual measurements of reconstructing in vivo dataset at $R = 4$ 2D Poisson disk under-sampling pattern in calibration-free parallel imaging by LINDBERG and EBMRec. The intensity of error maps is five times magnified. From top to bottom: The reconstruction results in 8 coils and 12 coils, respectively.

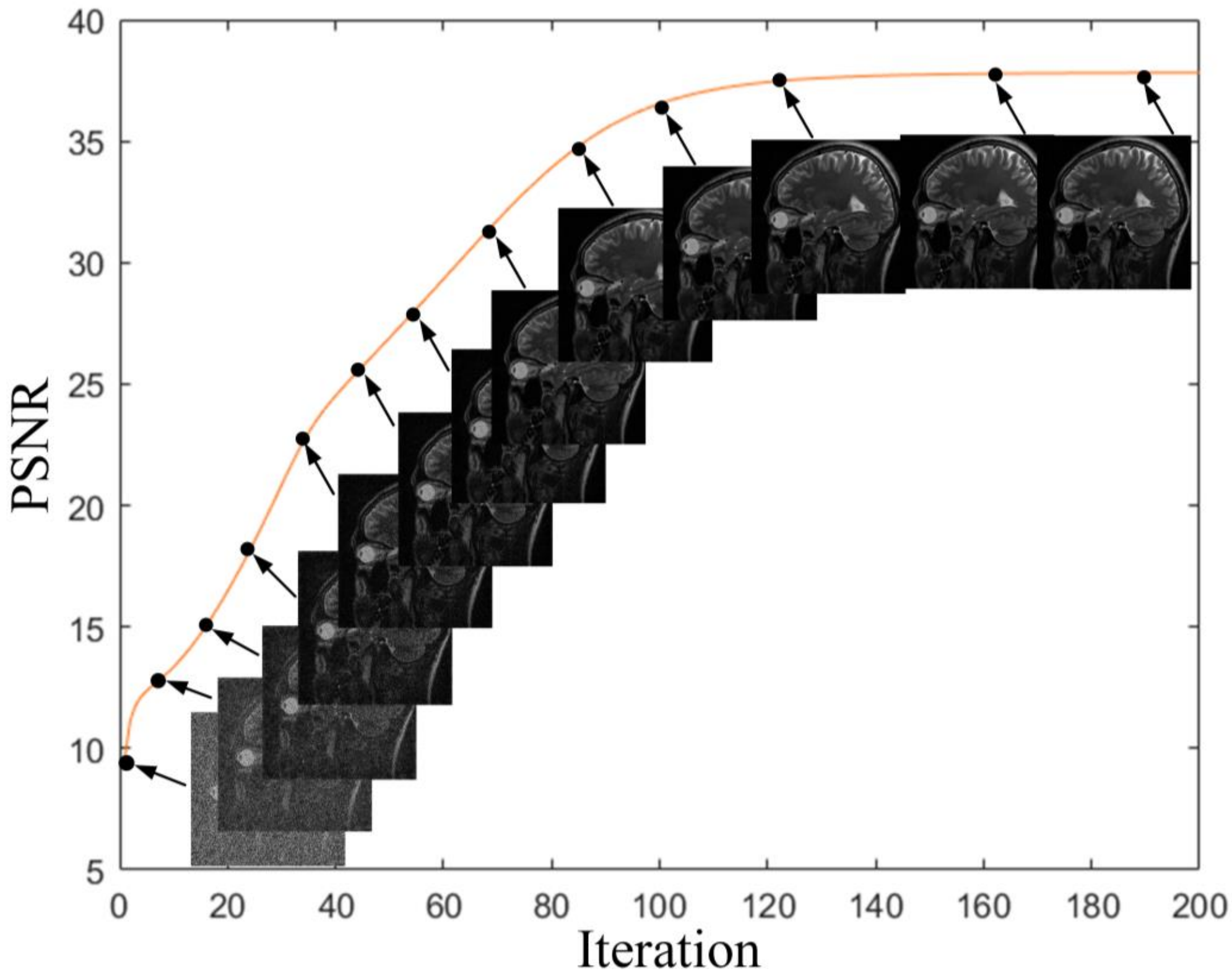

**FIG. 8.** Quantitative PSNR curve versus iteration for reconstructing the 5th brain image. The convergence tendency reflects the stability of iterative steps.

**Table 1.** PSNR and SSIM comparison of different methods for single coil reconstruction under different under-sampling patterns at a variety of acceleration factors. Among them, 31 and 50 test images are used for single coil reconstruction including brain and knee MRI.

| **Single coil reconstruction** | | | | | |
|---|---|---|---|---|---|
| Acceleration factor | Dataset | DLMRI | RefineGAN | EDAEPRec | EBMRec |
| *R*=3.3<br>Radial | *SIAT* | 33.43<br>0.905 | 34.91<br>0.938 | 35.62<br>0.928 | 35.84<br>0.925 |
| *R*=4<br>Radial | *SIAT* | 32.41<br>0.887 | 32.81<br>0.919 | 34.49<br>0.915 | 34.96<br>0.911 |
| *R*=5<br>Radial | *SIAT* | 31.21<br>0.860 | 30.52<br>0.872 | 33.49<br>0.899 | 33.80<br>0.892 |
| *R*=6.7<br>Random | *SIAT* | 27.63<br>0.752 | 25.95<br>0.721 | 30.68<br>0.843 | 30.97<br>0.852 |
| Acceleration factor | Dataset | LORAKS | U-Net | MEDMSPRec | EBMRec |
| *R*=4<br>Cartesian | *FastMRI* | 28.49<br>0.708 | 30.05<br>0.789 | 30.38<br>0.765 | 30.78<br>0.766 |
| *R*=8<br>Cartesian | *FastMRI* | 26.05<br>0.637 | 26.92<br>0.716 | 27.97<br>0.679 | 27.97<br>0.680 |

**Table 2.** Quantitative PSNR and SSIM measures for different algorithms with varying accelerate factors and sampling masks. Among them, DDP results of test images at $R=2$ and $R=3$ 1D Cartesian under-sampling pattern are obtained in 15 coils parallel imaging, while reconstruction MoDL results of 10 brain test images at $R=6$ 1D pseudo random sampling in 12 coils parallel imaging.

| **Parallel imaging reconstruction with knowing coil-sensitive** | | | | | |
|---|---|---|---|---|---|
| Acceleration factor | Dataset | Zero-filled | MoDL | DDP | EBMRec |
| *R*=2<br>Cartesian | *Test2* | 30.55<br>0.834 | -<br>- | 37.31<br>0.946 | 39.55<br>0.957 |
| *R*=3<br>Cartesian | *Test2* | 26.69<br>0.746 | -<br>- | 33.47<br>0.906 | 37.26<br>0.929 |
| *R*=6<br>Random | *Test1* | 25.75<br>0.774 | 39.65<br>0.936 | -<br>- | 42.20<br>0.989 |

**Table 3.** PSNR and SSIM comparison of in vivo dataset for calibration-free parallel imaging at 2D Poisson disk under-sampling with varying accelerate factors and different number of receiver coils. From top to bottom: the proposed method EBMRec and LINDBERG, respectively.

| **Calibration-free parallel imaging reconstruction** | | | | |
|---|---|---|---|---|
| Coil number | *R*=3 | *R*=4 | *R*=5 | *R*=6 |
| 4 | 44.02\0.983 | 41.99\0.976 | 39.21\0.969 | 37.15\0.961 |
| | 42.50\0.979 | 40.45\0.972 | 38.50\0.966 | 37.06\0.961 |
| 8 | 43.84\0.987 | 41.99\0.981 | 39.27\0.975 | 37.01\0.968 |
| | 42.38\0.972 | 40.43\0.962 | 38.28\0.953 | 36.66\0.946 |
| 12 | 33.46\0.903 | 31.85\0.889 | 30.21\0.864 | 29.15\0.843 |
| | 33.42\0.896 | 31.65\0.865 | 30.11\0.836 | 28.28\0.812 |

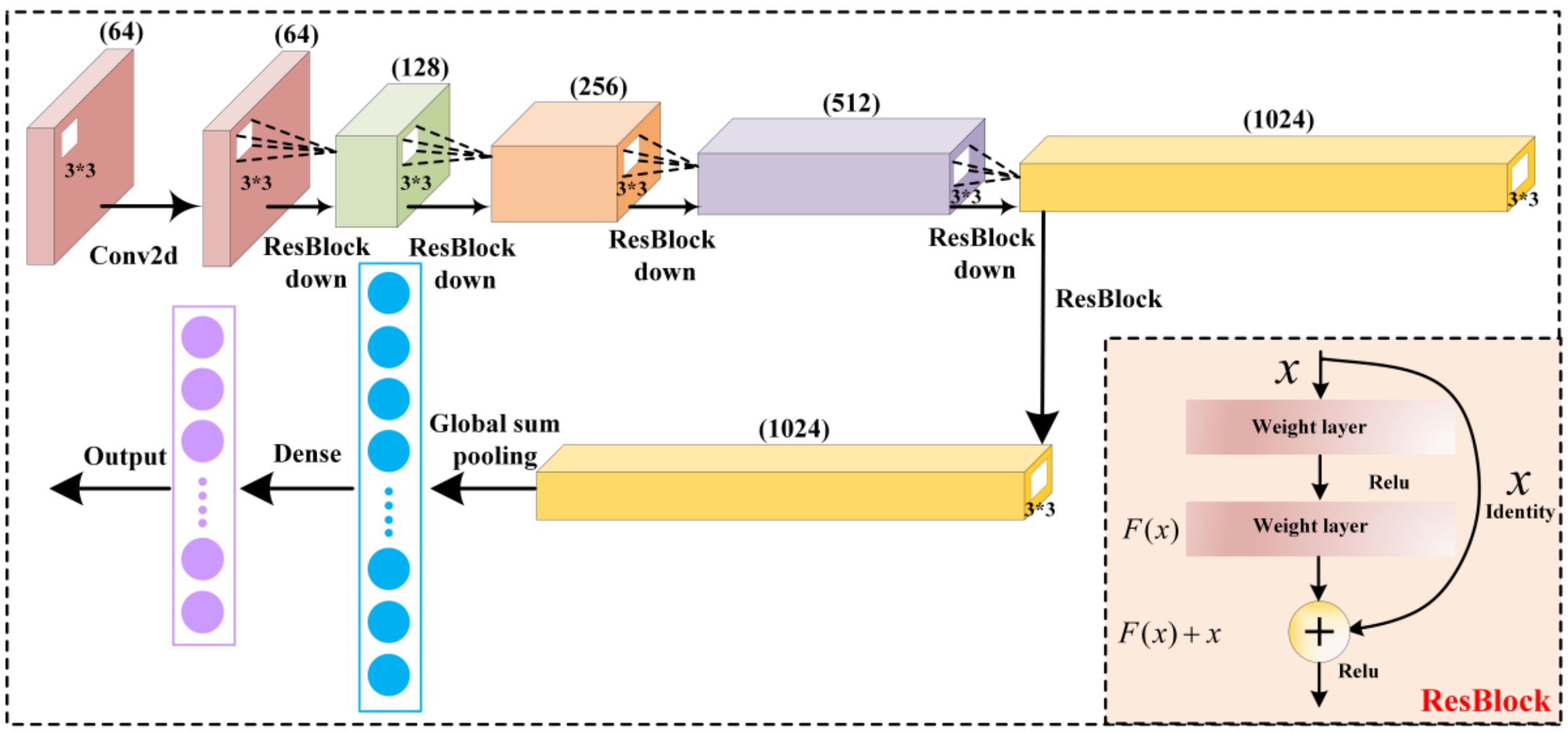

**Supporting Information Figure S1.** The diagram shows the residual model with multiple residual network architectures and parameters, which is the prior model used in this study.

**Supporting Information Table S1.** Computational costs of EBMRec and the competing methods. Notice that the reference image is a $256\times256$ MRI image. DLMRI is executed with CPU, while RefineGAN, EDAEPRec and EBMRec are with GPU.

| Runtime (s) | DLMRI | RefineGAN | EDAEPRec | EBMRec |
|---|---|---|---|---|
| Total time | 430.36 | 9.10 | 82.88 | 36.50 |
| Iter time | 17.21/Iter | --/-- | 0.42/Iter | 0.10/Iter |

**Supporting Information Table S2.** Performance of reconstructing 31 test images by different initial values at radial sampling trajectories in single coil.

| Initial value | | $R$=3.3 | $R$=4 | $R$=5 |
|---|---|---|---|---|
| Uniform noise $x \sim N(-1,1)$ | PSNR | 35.84 | 34.96 | 33.80 |
| | SSIM | 0.925 | 0.9114 | 0.892 |
| Zero-filled | PSNR | 35.85 | 34.96 | 33.81 |
| | SSIM | 0.926 | 0.9113 | 0.894 |